\documentclass[aps,preprintnumbers,prb,showpacs,superscriptaddress]{revtex4}
\usepackage{mathrsfs}
\usepackage{amsfonts}
\usepackage{graphicx}
\usepackage{amsmath}
\usepackage{amssymb}
\begin{document}
\title{Description of spin transport and precession in spin-orbit coupling systems and
general equation of continuity}
\author{Hua-Tong Yang}
\email{htyang@pku.edu.cn} \affiliation{School of Physics, Peking
University, Beijing 100871, China}
\author{Chengshi Liu} \affiliation{Department of
Mathematics, Daqing Petroleum Institute, Daqing 163318, China}

\begin{abstract}
By generalizing the usual current density to a matrix with respect
to spin variables, a general equation of continuity satisfied by the
density matrix and current density matrix has been derived. This
equation holds in arbitrary spin-orbit coupling systems as long as
its Hamiltonian can be expressed in terms of a power series in
momentum. Thereby, the expressions of the current density matrix and
a torque density matrix are obtained. The current density matrix
completely describes both the usual current and spin current as
well; while the torque density matrix describes the spin precession
caused by a total effective magnetic field, which may include a
realistic and an effective one due to the spin-orbit coupling. In
contrast to the conventional definition of spin current, this
expression contains an additional term if the Hamiltonian includes
nonlinear spin-orbit couplings. Moreover, if the degree of the full
Hamiltonian $\geq3$, then the particle current must also be modified
in order to satisfy the local conservation law of number.
\end{abstract}
\pacs{72.25.Dc, 72.25.Ba, 72.25.Mk}
\maketitle
\section{Introduction}
Recently, the spin transport, precession, and accumulation generated
by spin-orbit coupling and the resulting intrinsic spin-Hall
effect\cite{Zhang, Sinova} have attracted considerable interest by
physicists,\cite{Inoue,Dimitrova,Mishchenko,Schliemann&Loss-69,
Nomura, Chalaev and Loss, Raimondi and Schwab, Khaetskii,
Schliemann&Loss-71, Liu & Lei, Bryksin} because it may provide a
promising method to manipulate the electronic spin without
application of magnetic field\cite{Kato-Nature, Kato, Wunderlich,
Koga} in spintronics.\cite{Rashba-0309441, Zutic et al. rmp 76}
However, some conceptual ambiguities still remain in how to properly
describe the spin transport and precession in spin-orbit coupling
systems. In order to describe these phenomena, a conventional spin
current operator is intuitively defined as
$\hat{J}^{j}_{i}=(1/2)\{\hat{v}_i, \hat{s}_j\}$, with
$\hat{v}_i=\partial \hat{H}/\partial \hat{p}_i$ a velocity operator
and $\hat{s}_j$ is a spin component (actually, as will be pointed
out in Sec. IV, this expression should also contain a factor of
density operator $\delta(\hat{\bf x}-{\bf x})$ in order to have the
dimension of a current density). However, it is still very
questionable whether this quantity is meaningfully observable. This
problem has caused a lot of confusion and debates in recent
literature\cite{Rashba-0311110,Sun & Xie,Shi,Lin-Li-zhang, YongWang,
ShenR} and has even become an obstacle to the further progress of
this filed. From the experimental point of view, the most
interesting observable is the spin
density,\cite{Kato,Wunderlich,Koga} which can be given by
$\textmd{tr}(\hat{s}_j \hat{\rho})$, with
$$\hat{\rho}\equiv\left[\begin{array}{cc}\rho_{\uparrow\uparrow}&
\rho_{\uparrow\downarrow}\\ \rho_{\downarrow\uparrow}&
\rho_{\downarrow\downarrow}\end{array}\right]$$ the density matrix;
i.e., the density matrix $\hat{\rho}$ can completely describe the
number density as well as its spin polarization. Theoretically, a
fundamental equation describing a transport phenomenon of any
additive conserved quantity, say $Q$, is the equation of continuity
$\partial \rho_Q/\partial t+\nabla \cdot{\bf j}_Q=0$, which connects
the change of its density $\rho_Q$ to a corresponding current ${\bf
j}_Q$; if this additive quantity is not conserved, we must introduce
a source (or drain) term $S_Q$ in this equation such that $\partial
\rho_Q/\partial t+\nabla \cdot{\bf j}_Q=S_Q$. As a classical picture
of motion, the spin transport in spin-orbit coupling systems is very
similar to the latter case, since the spin is also an additive and
local quantity (the locality means that the spin commutes with the
position); i.e., the spin density can be defined as a function of
position. Therefore, a straightforward approach to introduce the
spin current is to establish a continuitylike equation satisfied by
the density matrix $\hat{\rho}$. This equation should be a general
counterpart of the usual continuity equation of number $\partial
\rho/\partial t+\nabla \cdot{\bf J}=0$, and can be called a general
continuity equation. We expect that a corresponding current density
matrix $\hat{\bf J}$ can be naturally obtained from it. However, as
the spin is not conserved, so the time rate of change of
$\hat{\rho}$ must contain two distinct parts, i.e., $\partial
\hat{\rho}/\partial t=-\nabla \cdot\hat{\bf J}+\hat{T}$. The first
part is a divergence $-\nabla \cdot\hat{\bf J}$, which can be
ascribed to a current density matrix $\hat{\bf J}$; while the second
one, denoted by $\hat{T}$, does not possess this character and can
be called a torque density matrix. If we integrate both sides of
this equation over a volume $\Omega$, then we have
$(\partial/\partial t)\int
_{\Omega}\hat{\rho}dV=-\oint_{\partial\Omega} d{\bf S}\cdot\hat{\bf
J}+\int_{\Omega}\hat{T}dV$. The surface integral
$\oint_{\partial\Omega} d{\bf S}\cdot\hat{\bf J}$ can be interpreted
as a current flowing through the boundary surface $\partial\Omega$,
while the volume integral $\int_{\Omega}\hat{T}dV$ represents the
precession rate of the spin polarization within the volume $\Omega$.
As will be demonstrated in Sec. II, this division can be uniquely
determined by the equation of motion $\partial \hat{\rho}/\partial
t=[\hat{H}({\bf p},{\bf x}),
\hat{\rho}]\equiv\mathcal{L}[\hat{\rho}({\bf x},{\bf x}';t)]$ and
the condition that $\hat{T}$ can no longer be expressed as a
divergence, where
$\mathcal{L}[\hat{\rho}]=\mathcal{L}(\nabla,\nabla')[\hat{\rho}]$ is
a functional operator acting on $\hat{\rho}$ and $\nabla=\nabla_{\bf
x}$ and $\nabla'=\nabla_{\bf x'}$ are the differentiation operators
with respect to ${\bf x}$ and ${\bf x}'$, respectively. The essence
of this problem is to cast the $\mathcal{L}[\hat{\rho}]$ into the
form
$$\mathcal{L}(\nabla,\nabla')[\hat{\rho}]=(\nabla+\nabla')\cdot\vec{\mathcal{J}}[\hat{\rho}]
+\mathcal{T}(\frac{1}{2}(\nabla-\nabla'))[\hat{\rho}],$$ where
$\vec{\mathcal{J}}(\nabla,\nabla')[\hat{\rho}]$ and
$\mathcal{T}(\frac{1}{2}(\nabla-\nabla'))[\hat{\rho}]$ are also
functional operators. It will be proven that $\vec{\mathcal{J}}$ and
$\mathcal{T}$ can be uniquely determined by the equation of motion
as long as $\mathcal{L}$ can be written as a power series in
$\nabla$ and $\nabla'$. So, if we define
$$\hat{\bf J}({\bf x})=-\lim_{{\bf x}'\rightarrow{\bf
x}}\vec{\mathcal{J}}(\nabla,\nabla')[\hat{\rho}({\bf x},{\bf
x}')],$$ $$\hat{T}({\bf x})=\lim_{{\bf x}'\rightarrow{\bf x}}
\mathcal{T}(\frac{1}{2}(\nabla-\nabla'))[\hat{\rho}({\bf x},{\bf
x}')],$$ then the continuity-like equation can be obtained.  It is
important to note that if we transform the $\hat{\rho}({\bf x},{\bf
x}')$ to the Wigner function $\hat{\rho}(\tilde{\bf p},{\bf R})$,
where ${\bf R}=\frac{1}{2}({\bf x}+{\bf x}')$ and $\tilde{\bf p}$ is
the canonical momentum corresponding to ${\bf x}-{\bf x}'$, then the
expression for $\hat{T}$ has the form $\hat{T}({\bf R})=\int
d^3\tilde{\bf p}\mathcal{T}(i\tilde{\bf p})[\hat{\rho}(\tilde{\bf
p},{\bf R})]$; it no longer contains the spatial derivative
$\nabla_{\bf R}$ and, therefore, can not be written as a divergence.

From this general continuity equation, we can deduce that the
$\textmd{tr}\hat{\bf J}$ and $\textmd{tr}(\hat{\bf s}\hat{\bf J})$
represent the current density of particle ${\bf J}$ and the current
density of spin ${\bf J}^{\bf s}$, respectively, just as the
$\textmd{tr}\hat{\rho}$ and $\textmd{tr}(\hat{\bf s}\hat{\rho})$
represent the number density $\rho$ and spin density $\rho_{\bf
s}({\bf x})$, respectively, i.e., the matrix $\hat{\bf J}$ can
provide a complete description of the particle transport as well as
the spin transport. In order to confirm that this physical
interpretation is self-consistent, we must prove that (1)
$\textmd{tr}\hat{T}\equiv0$, which is a necessary and sufficient
condition for the local conservation of number and (2) $\hat{\bf J}$
and $\hat{T}$ must be gauge invariant in order to ensure that they
can be interpreted as physically observable. In this paper, a
general equation of continuity satisfied by $\hat{\rho}$, $\hat{\bf
J}$ and $\hat{T}$ is derived, which holds in arbitrary spin-orbit
coupling systems as long as its Hamiltonian can be written as a
polynomial in momentum. Consequently, the expressions of the
$\hat{\bf J}$ and $\hat{T}$ are presented. These expressions are
gauge invariant and independent of the magnitude of spin and the
system dimension and satisfy $\textmd{tr}\hat{T}\equiv0$. In
contrast to the conventional definition, this formulism contains the
following corrections: First, if the spin-orbit coupling Hamiltonian
includes second- or higher order power of momentum, e.g., in the
Luttinger model,\cite{Luttinger, Zhang} the two-dimensional (2D)
cubic Rashba model,\cite{Schliemann&Loss-71} or the Dresselhaus
spin-orbit coupling,\cite{Dresselhaus, Rashba-Sov,Dyakonov} then we
should add a new additional term to the expression for $\hat{\bf
J}$, which is purely originated from the nonlinear spin-orbit
coupling and contributes only to the spin current. Second, if the
degree of the full Hamiltonian $\textmd{deg} \hat{H}(\hat{\bf
p})\geq 3$, e.g., in the cubic Rashba model or Dresselhaus
spin-orbit term, then the particle current density ${\bf J}$ as well
as the spin current density are also different from the conventional
definition; otherwise, it will violate the local conservation law of
number. Moreover, this equation of continuity also holds in the
presence of arbitrary electromagnetic field and Coulomb interaction
or nonmagnetic impurity scattering. Consequently, from this
continuity-like equation we can also deduce some exact identities
which are very similar to the Ward-Takahashi identity, although the
spin is not a conserved quantity.

This paper is organized as follows. In Sec. II the general equation
of continuity in a spin-orbit coupling system without Coulomb
interaction and impurity scattering is derived. Then, we will prove
that this equation also holds in the presence of the Coulomb
interaction and nonmagnetic impurities in Sec. III. Section IV is
devoted to a detail discussion and comparison of these expressions
with the conventional intuitive definition. A short summary of the
conclusions will be given in Sec. V.
\section{General equation of continuity}
Firstly, we consider a general spin-orbit coupling system without
any two-body interaction or impurity scattering, but may subject to
an external electromagnetic field described by $(\phi, {\bf A} )$;
the case in the presence of the Coulomb interaction or impurity
scattering will be discussed in Sec. III. The corresponding one-body
Hamiltonian has the form
\begin{eqnarray}
\hat{H}&=&\frac{1}{2m}\left({\bf P}-\frac{e}{c}{\bf A}\right)^2
+e\phi+\hat{h}_{\textmd{so}} =
\hat{h}_0+e\phi+\hat{h}_{\textmd{so}}, \label{Hamiltonian}
\end{eqnarray}
where $\hat{h}_0\equiv(1/2m)({\bf P}-\frac{e}{c}{\bf A})^2$ is the
kinetic energy and $\hat{h}_{\textmd{so}}$ is a general spin-orbit
coupling Hamiltonian and we suppose it can be expanded as a power
series
\begin{eqnarray}
\hat{h}_{\textmd{so}}=\hat{H}^{(0)}+\hat{H}^{(1)}_{i}\hat{p}_{i}+\hat{H}^{(2)}_{ij}
\hat{p}_{i}\hat{p}_{j}+\hat{H}^{(3)}_{ijk}\hat{p}_{i}\hat{p}_{j}\hat{p}_{k}+\cdots.\label{so-coupling}
\end{eqnarray}
Here and hereafter, a summation over repeated suffix is implied,
where $\hat{p}_i\equiv -i\hbar{\partial}_i-(e/c)A_{i}\equiv
-i\hbar\widetilde{\partial}_i$ $(i=x,y,z),$ $\widetilde{\partial}_i$
are the covariant derivatives, and $\hat{H}^{(n)}_{i_1\cdots i_n}$
$(n=0,1,2,\cdots)$ are the operators acting on spin variable (or
$2\times 2$ matrixes if the spin is $1/2$); meanwhile,
$\hat{H}^{(0)}$ is a scalar representing Zeeman energy due to
magnetic field, $\hat{H}^{(1)}_i$ are the components of a vector,
$\hat{H}^{(n)}_{i_{1}\cdots i_{n}}$ are the components of a tensor
of rank $n$th and are supposed to be symmetric under the
permutations of its suffixes $i_1,i_2,\ldots, i_n$. Furthermore, we
assume that all $\hat{H}^{(n)}_{i_1\cdots i_n}$ are independent of
position except $\hat{H}^{(0)}$. The corresponding many-body
Hamiltonian is $\hat{\bf H}=\int
\psi^{\dag}_{\sigma}(1)\hat{H}_{\sigma\sigma'}\psi_{\sigma'}(1)$,
where $\sigma$ and $\sigma'$ are spin variables.

In order to derive the general equation of continuity, we consider
the equation of motion of Green's function, for this approach is
apparently gauge invariant. The density matrix $\hat{\rho}$ can be
given by $\rho_{\alpha\beta}({\bf x}t)=-\langle
\hat{\mathscr{T}}[\psi_{\alpha}({\bf x}t)\psi_{\beta}^\dag({\bf
x}t^+)] \rangle=-i\hbar G_{\alpha\beta}({\bf x}t,{\bf x}t^+),$ where
$\hat{\mathscr{T}}$ is the time-ordering operator and $t^+\equiv
t+0^+$. From the equation of motion we have\cite{K&B-book,
Haug&Jauho}
\begin{eqnarray}
\left[\left(i\hbar\frac{\partial}{\partial t}-e\phi(1)\right
)-\left(-i\hbar\frac{\partial}{\partial t'}-e\phi(1')\right )\right
]\hat{G}(1,1')-\left[\hat{h}_0+\hat{h}_{\textmd{so}},\hat{G}(1,1')\right
]=0,\label{subtract-Dyson}
\end{eqnarray}
where $ \hat{G}$ denotes the matrix $G_{\alpha\beta}$, and
$(1)\equiv({\bf x}t),(1')\equiv({\bf x}'t')$. By taking the limit as
$1'\rightarrow 1^+$, the first term becomes
\begin{eqnarray}\lim_{1'\rightarrow 1^+}
\left [ i\hbar\frac{\partial}{\partial t}-e\phi(1)
+i\hbar\frac{\partial}{\partial t'}+e\phi(1')\right ]\hat{G}(1,1')=
-\frac{\partial \hat{\rho}({\bf x}t)}{\partial
t}.\label{time-differ}
\end{eqnarray}
The commutator $[h_0,\hat{G}]$ in the second term of Eq.
(\ref{subtract-Dyson}) can be written as \begin{eqnarray} \left
[h_0, \hat{G}\right ]=\left [{\bf P}-\frac{e}{c}{\bf
A},\frac{1}{2m}\left \{{\bf P}-\frac{e}{c}{\bf
A},\hat{G}(1,1')\right \}\right], \nonumber\end{eqnarray} where
$({\bf P}-(e/c){\bf A})\hat{G}(1,1')\equiv[-i\hbar\nabla-(e/c){\bf
A}(1)]\hat{G}(1,1'),$ $\hat{G}(1,1')({\bf P}-(e/c){\bf
A})\equiv[i\hbar\nabla'-(e/c){\bf A}(1')]\hat{G}(1,1'),$ $[\hat{\bf
X},\hat{\bf Y}]\equiv\sum_i[\hat{X}_i, \hat{Y}_i],$ and $\{\hat{\bf
X},\hat{ Z}\}\equiv\sum_i\{\hat{X}_i, \hat{Z}\}.$ Because
$$ \lim_{1'\rightarrow 1^+}\left \{{\bf P}-\frac{e}{c}{\bf A} ,~
\hat{G}(1,1')\right \}=\frac{i}{\hbar} \left \{{\bf
P}-\frac{e}{c}{\bf A},~ \hat{\rho}(1,1')\right \}_{1'=1},$$ so, if
we define a current density matrix
\begin{eqnarray}\hat{\bf
J}_0(1)\equiv\frac{1}{2m}\left \{{\bf P}-\frac{e}{c}{\bf A},~
\hat{\rho}(1,1')\right \}_{1'=1},\label{current-j0}\end{eqnarray}
then we have
\begin{eqnarray}\lim_{1'\rightarrow 1^+}\left [\hat{h}_0, \hat{G}(1,1')\right ]
=\nabla\cdot \hat{\bf J}_0({\bf
x}t).\label{divergence-0}\end{eqnarray} Now, we consider the
spin-orbit coupling term $[\hat{h}_{\textmd{so}},\hat{G}(1,1')]$.
According to Eq. (\ref{so-coupling}), the first term is simply
\begin{eqnarray}\lim_{1'\rightarrow 1^+}\left[\hat{H}^{(0)},
\hat{G}\right]=-\frac{1}{i\hbar}\left[\hat{H}^{(0)},
\hat{\rho}\right]_{1'=1}.\label{torque-0}\end{eqnarray} In order to
treat the other terms of $[\hat{h}_{\textmd{so}},\hat{G}(1,1')]$, we
note that $\hat{H}^{(n)}_{i_1\cdots i_n}$ $(n\geq1)$ commute with
$\hat{p}_{i}$ because they are independent of the position, and use
the following operator relation: if $[\hat{A},\hat{B}]=0,$ then
$[\hat{A}\hat{B},
\hat{C}]=(1/2)\{\hat{A},[\hat{B},\hat{C}]\}+(1/2)[\hat{A},\{\hat{B},\hat{C}\}].$
By substituting Eq. (\ref{so-coupling}) into
$[\hat{h}_{\textmd{so}},\hat{G}(1,1')]$, we obtain
\begin{eqnarray}\left[\hat{h}_{\textmd{so}}, \hat{G}\right]=\frac{1}{2}\sum_{n}\left(\left
\{\hat{H}^{(n)}_{i_1\cdots i_n},\left[ \hat{p}_{i_1}\cdots
\hat{p}_{i_n}, \hat{G} \right ]\right \}+\left
[\hat{H}^{(n)}_{i_1\cdots i_n},\left \{ \hat{p}_{i_1}\cdots
\hat{p}_{i_n}, \hat{G} \right \}\right ]\right
),\label{SO-coupling}\end{eqnarray} where
\begin{eqnarray}&&\left[ \hat{p}_{i_1}\cdots \hat{p}_{i_n}, \hat{G} \right
]\equiv\left[\hat{p}_{i_1}\cdots \hat{p}_{i_n}-\hat{p}'_{i_1}\cdots
\hat{p}'_{i_n}\right]\hat{G}(1,1'),\label{commutator-1}
\end{eqnarray}
\begin{eqnarray}
&&\left\{ \hat{p}_{i_1}\cdots
\hat{p}_{i_n},\hat{G}\right\}\equiv\left(\hat{p}_{i_1}\cdots
\hat{p}_{i_n}+\hat{p}'_{i_1}\cdots
\hat{p}'_{i_n}\right)\hat{G}(1,1'),\label{anticommutator}
\end{eqnarray} and $\hat{p}'_{j}\equiv i\hbar\partial'_{j}-(e/c) A_j(1')\equiv
i\hbar\widetilde{\partial}'_{j},$ which commute with $\hat{p}_i$
because $\hat{p}_i$ act on $1=({\bf x},t)$ while $\hat{p}'_{j}$ act
on $1'=({\bf x}',t')$. It is easy to verify that Eq.
(\ref{torque-0}) is also a special case of identity
(\ref{SO-coupling}). The limit of the first term on the right-hand
side of Eq.(\ref{SO-coupling}) can be entirely written as a
divergence form. For $n=1$, we have
\begin{eqnarray}\lim_{1'\rightarrow 1^+}\frac{1}{2}\{H^{(1)}_i,
[\hat{p}_i,\hat{G}]\}=\nabla \cdot \frac{1}{2}\left\{{\bf H}^{(1)},
\hat{\rho}(1,1')\right\}_{1'=1},\label{current-n1}\end{eqnarray}
where ${\bf
H}^{(1)}=(\hat{H}^{(1)}_x,\hat{H}^{(1)}_y,\hat{H}^{(1)}_z)$. For
$n\equiv m+1\geq2$, we define an auxiliary operator
\begin{eqnarray}
\hat{D}^{(m)}_{i_1 \cdots
i_{m}}(\hat{p}_i,\hat{p}'_{i})\equiv\hat{p}_{i_1}\cdots
\hat{p}_{i_{m}}+\hat{p}_{i_1}\cdots
\hat{p}_{i_{m-1}}\hat{p}'_{i_{m}}+\cdots +\hat{p}'_{i_1}\cdots
\hat{p}'_{i_{m}}=\sum^{m}_{l=0}\hat{p}_{i_1}\cdots
\hat{p}_{i_{m-l}}\hat{p}'_{i_{m-l+1}}\cdots\hat{p}'_{i_{m}}.
\label{D-operator}
\end{eqnarray}
Owing to the symmetry of $\hat{H}^{(n)}_{i_1 i_2\cdots i_n}$ and
commutability between $\hat{p}_i$ and $\hat{p}'_{j},$ we have
\cite{Takahashi}
\begin{eqnarray}\frac{1}{2}\left\{\hat{H}^{(m+1)}_{i_1 i_2\cdots i_{m+1}}
, \left[\hat{p}_{i_1}\hat{p}_{i_2}\cdots
\hat{p}_{i_{m+1}},\hat{G}(1,1')\right]\right\}=(\hat{p}_i-\hat{p}'_{i})\frac{1}{2}\left\{\hat{H}^{(m+1)}_{i,
i_1\cdots i_{m}},\hat{D}^{(m)}_{i_1 \cdots
i_{m}}\hat{G}(1,1')\right\}.\label{Takahashi-identity}\end{eqnarray}
So, if we define
\begin{eqnarray}(\hat{J}_{\textmd{so}})_{i}\equiv\sum_{m}
\frac{1}{2}\left\{\hat{H}^{(m+1)}_{i, i_1\cdots
i_{m}},\hat{D}^{(m)}_{i_1\cdots i_{m}}\hat{\rho}(1,1')\right
\}_{1'=1},\label{current_so}\end{eqnarray} where we also introduce a
$\hat{D}^{(0)}\equiv1$ in order to include the term of Eq.
(\ref{current-n1}) in this unified formula, then we have
\begin{eqnarray}\lim_{1'\rightarrow1^+}\sum_{n}\frac{1}{2}\left\{\hat{H}^{(n)}_{i_1\cdots i_n}
, \left[\hat{p}_{i_1}\cdots
\hat{p}_{i_n},\hat{G}(1,1')\right]\right\}=\partial_i
(\hat{J}_{\textmd{so}})_{i}=\nabla\cdot \hat{\bf
J}_{\textmd{so}}.\label{divergenec-1}
\end{eqnarray}
The limit of the second term $\frac{1}{2}\left [\hat{H}^{(n)}_{i_1
\cdots i_n},\left \{ \hat{p}_{i_1}\cdots \hat{p}_{i_n} \hat{G}
\right \} \right ]$ on the righ-hand side of Eq. (\ref{SO-coupling})
cannot be entirely written as a divergence. However, we can prove
that it can be written as a divergence and a remaining term, while
the remaining term no longer contains spatial derivatives
$\nabla_{\bf R}$ if it is represented by the Wigner distribution
function. In order to prove this conclusion, we take first the case
when $n=1$, it evidently cannot be written as a divergence as
$1'\rightarrow 1^+$, because $\{\hat{p}_i,
\hat{G}\}=-i\hbar(\widetilde{\partial}_i-\widetilde{\partial}'_{i})\hat{G}(1,1')$
and only the limit of
$(\widetilde{\partial}_i+\widetilde{\partial}'_{i})$ can be written
as a divergence. Then, we take $n\equiv m+1\geq 2$ and prove that
the limits of
\begin{eqnarray}\hat{H}^{(n)}_{i_1 i_2\cdots i_n}\left \{
\hat{p}_{i_1}\hat{p}_{i_2}\cdots \hat{p}_{i_n}, \hat{G}\right \} -2
\hat{H}^{(n)}_{i_1 i_2\cdots
i_n}\frac{\hat{p}_{i_1}+\hat{p}'_{i_1}}{2}
\frac{\hat{p}_{i_2}+\hat{p}'_{i_2}}{2}\cdots
\frac{\hat{p}_{i_n}+\hat{p}'_{i_n}}{2}
\hat{G}\nonumber\end{eqnarray} and
\begin{eqnarray}\left \{
\hat{p}_{i_1}\hat{p}_{i_2}\cdots \hat{p}_{i_n}, \hat{G}\right
\}\hat{H}^{(n)}_{i_1 i_2\cdots i_n} -2
\frac{\hat{p}_{i_1}+\hat{p}'_{i_1}}{2}
\frac{\hat{p}_{i_2}+\hat{p}'_{i_2}}{2}\cdots
\frac{\hat{p}_{i_n}+\hat{p}'_{i_n}}{2} \hat{G}\hat{H}^{(n)}_{i_1
i_2\cdots i_n}\nonumber\end{eqnarray} can be written as a divergence
form. To this end, we again introduce the following auxiliary
operators:
\begin{eqnarray}\hat{R}^{(n)}_{i_1 \cdots i_{n}}(\hat{p}_i,\hat{p}'_{i})\equiv\frac{\hat{p}_{i_1}+\hat{p}'_{i_1}}{2}
\frac{\hat{p}_{i_2}+\hat{p}'_{i_2}}{2}\cdots
\frac{\hat{p}_{i_n}+\hat{p}'_{i_n}}{2},\end{eqnarray}
\begin{eqnarray}\hat{C}^{(m)}_{i_1 \cdots i_{m}}(\hat{p}_i,\hat{p}'_{i})&\equiv&\hat{p}_{i_1}\cdots
\hat{p}_{i_{m}}+\hat{p}_{i_1}\cdots\hat{p}_{i_{m-1}}\hat{R}^{(1)}_{i_{m}}\cdots
+\hat{p}_{i_1}\hat{R}^{(m-1)}_{i_2\cdots i_{m}}\nonumber\\&-&
\hat{p}'_{i_1}\cdots
\hat{p}'_{i_{m}}-\hat{p}'_{i_1}\cdots\hat{p}'_{i_{m-1}}\hat{R}^{(1)}_{i_{m}}
\cdots -\hat{p}'_{i_1}\hat{R}^{(m-1)}_{i_2\cdots
i_{m}},\label{C-operator}
\end{eqnarray}
and define another term of the current density matrix $\hat{\bf
J}'_{\textmd{so}}$ as
\begin{eqnarray}{(\hat{J}'_{\textmd{so}})_i}\equiv\sum_{m\geq 1}\frac{1}{4}
\left[\hat{H}^{(m+1)}_{i,i_1\cdots i_{m}},\hat{C}^{(m)}_{i_1\cdots
i_{m}}\hat{\rho}(1,1')\right ]_{1'=1}\label{current-1}
\end{eqnarray} and a
torque density matrix as
\begin{eqnarray}\hat{T}[\hat{\rho}]\equiv\sum_{n}
\frac{1}{i\hbar}\left[\hat{H}^{(n)}_{i_1i_2\cdots
i_{n}},\hat{R}^{(n)}_{i_1 i_2 \cdots i_{n}}\hat{\rho}(1,1')\right
]_{1'=1}.\label{torque}\end{eqnarray} It is easy to verify that the
terms of $n=0, 1$ can also be incorporated into the righ-hand side
of definition (\ref{torque}) if we define $\hat{R}^{(0)}\equiv1$.
Then, we can prove that (see the Appendix A)
\begin{eqnarray}\lim_{1'\rightarrow1^+}\sum_{n}\frac{1}{2}\left [\hat{H}^{(n)}_{i_1
i_2\cdots i_n},\left \{ \hat{p}_{i_1}\hat{p}_{i_2}\cdots
\hat{p}_{i_n}, \hat{G} \right \} \right ]= \nabla\cdot\hat{\bf
J}'_{\textmd{so}}-\hat{T}[\hat{\rho}],\label{deverg-torque}\end{eqnarray}
where $\hat{\bf J}'_{\textmd{so}}$ is a new term of the current
density matrix, which is originated from the nonlinear spin-orbit
coupling, because it will vanish if $\hat{H}^{(n)}_{i_1 i_2\cdots
i_n}$ are independent of spin (i.e., it is only a c-number rather
than a spin matrix). More importantly, we should note that the
$\hat{T}[\hat{\rho}]$ cannot be rewritten as a divergence because it
is given by a polynomial of
$\widetilde{\partial}_i-\widetilde{\partial}'_i$. This property will
be more evident if we transform the $\hat{\rho}(1,1')$ to a
gauge-invariant Wigner distribution function\cite{Mahan} by
$$\hat{\rho}_{\alpha\beta}(\tilde{\bf p},\tilde{\varepsilon};{\bf
R},t)=\int d^3{\bf r}d\tau e^{
\{\frac{i}{\hbar}[(\widetilde{\varepsilon}+e\phi({\bf
R}t)]\tau-[\tilde{\bf p}+\frac{e}{c}{\bf A}({\bf R}t)]\cdot{\bf
r}\}}\langle \psi_{\beta}^\dag({\bf R}-\frac{\bf
r}{2},t-\frac{\tau}{2})\psi_{\alpha}({\bf R}+\frac{\bf
r}{2},t+\frac{\tau}{2})\rangle,$$ in which the macroscopic variables
${\bf R}=({\bf x}+{\bf x}')/2$ and microscopic variables ${\bf
r}={\bf x}-{\bf x}'$ have been used, and then we have
\begin{eqnarray}\left[\hat{H}^{(n)}_{i_1 \cdots i_n}, \hat{R}^{(n)}_{i_1 \cdots
i_{n}}\hat{\rho} \right]_{1'=1}=\int\frac{d^3\tilde{\bf
p}d\tilde{\varepsilon}}{(2\pi\hbar)^4}\left[\hat{H}^{(n)}_{i_1\cdots
i_n} \tilde{p}_{i_1}\cdots\tilde{p}_{i_n},\hat{\rho}(\tilde{\bf
p},\tilde{\varepsilon};{\bf
R};t)\right].\label{torque-p}\end{eqnarray} It does not involve the
macroscopic spatial derivative $\nabla_{\bf R}$, and therefore can
not be written as a divergence $\nabla_{\bf R}\cdot{\bf J}$.
Comparing this expression with Eq. (\ref{torque-0}), it is obvious
that the $\hat{H}^{(n)}_{i_1\cdots i_n}
\tilde{p}_{i_1}\cdots\tilde{p}_{i_n}$ plays a role which is very
similar to Zeeman energy $\hat{H}^{(0)}$. The only difference is
that the former depends on momentum, because it is originated from
the spin-orbit coupling, while the latter depends on position. So,
the $\hat{H}^{(n)}_{i_1\cdots i_n}
\tilde{p}_{i_1}\cdots\tilde{p}_{i_n}$ can be regarded as the energy
contributed by an effective magnetic filed depending on momentum.
Take the limit as $1'\rightarrow1^+$ and substituting Eq.
(\ref{time-differ}), (\ref{divergence-0}), (\ref{divergenec-1}) and
(\ref{deverg-torque}) into Eq. (\ref{subtract-Dyson}), we finally
obtain the following generalized equation of continuity:
\begin{eqnarray}
\frac{\partial \hat{\rho}}{\partial t}+\nabla \cdot \left (\hat{\bf
J}_0+\hat{\bf J}_{\textmd{so}}+\hat{\bf J}'_{\textmd{so}}\right
)=\hat{T}[\hat{\rho}].\label{continuity}
\end{eqnarray}
Evidently, the form of this equation is independent of the system
dimension or the magnitude of spin. The only differences are the
number of components of the current density and the dimension of the
matrices, if the system dimension or the spin magnitude are
different.

From the definitions (\ref{current-j0}),
(\ref{current_so}),(\ref{current-1}), and (\ref{torque}), we can
deduce that $\hat{\bf J}$ and $\hat{T}$ are gauge invariant because
all the derivatives in their definitions are covariant.\cite{Mahan,
Levanda} Moreover, from the definitions (\ref{current-1}) and
(\ref{torque}), we get $\textmd{tr}{\bf
\hat{J}}'_{\textmd{so}}\equiv0$ and
$\textmd{tr}(\hat{T}[\hat{\rho}])\equiv0$, owing to their commutator
form. The identity $\textmd{tr}(\hat{T}[\hat{\rho}])\equiv0$ ensures
the local conservation law of number (or charge), while
$\textmd{tr}{\bf \hat{J}}'_{\textmd{so}}\equiv0$ means that the
$\hat{\bf J}'_{\textmd{so}}$ does not contribute to the particle or
charge current, but it may contribute to the spin current. Because
$\textmd{tr}\hat{\rho}({\bf x}t)$ is the number density $\rho({\bf
x}t)$, so, $\textmd{tr} \hat{\bf J}({\bf x}t)=\textmd{tr}(\hat{\bf
J}_0+\hat{\bf J}_{\textmd{so}})$ is the current density of the
particle ${\bf J}({\bf x}t)$, while the charge current is ${\bf
J}_e({\bf x}t)=e{\bf J}({\bf x}t)$, here the identity
$\textmd{tr}{\bf \hat{J}}'_{\textmd{so}}\equiv0$ has been used. By
letting ${\bf J}_0=\textmd{tr}\hat{\bf J}_0$ and ${\bf
J}_{\textmd{so}}=\textmd{tr}\hat{\bf J}_{\textmd{so}}$ and taking
the trace over spin of both sides of the Eq. (\ref{continuity}), we
obtain the conservation law
\begin{eqnarray}
\frac{\partial \rho}{\partial t}+\nabla \cdot \left ({\bf J}_0+{\bf
J}_{\textmd{so}}\right )=0.
\end{eqnarray}
Similarly, because
\begin{eqnarray}\rho_{\bf s}({{\bf x}t})\equiv\textmd{tr}[\hat{\rho}({\bf
x}t)\hat{\bf s}] =\textmd{tr}\left[\hat{\bf s}\hat{\rho}({\bf
x}t)\right]=\frac{1}{2}\textmd{tr}\left\{\hat{\rho},\hat{\bf
s}\right\}\label{spin density}\end{eqnarray} is the expectation
value of the spin density, therefore, it is natural to define the
spin current density as
\begin{eqnarray}{\bf J}^{\bf s}({{\bf x}t})\equiv\textmd{tr}\left[\hat{\bf J}({\bf
x}t)\hat{\bf s}\right] =\textmd{tr}\left[\hat{\bf s}\hat{\bf J}({\bf
x}t)\right]=\frac{1}{2}\textmd{tr}\left\{\hat{\bf J},\hat{\bf
s}\right\}\label{spin-current}\end{eqnarray} and the spin torque
density as
\begin{eqnarray}T^{\bf s}({{\bf x}t})\equiv\textmd{tr}\left[\hat{T}({\bf x}t)\hat{\bf s}\right]=
\textmd{tr}\left[\hat{\bf s}\hat{T}({\bf
x}t)\right]=\frac{1}{2}\textmd{tr}\left\{\hat{T},\hat{\bf
s}\right\}.\label{torque-s}\end{eqnarray} From Eq.
(\ref{continuity}), we obtain the following equation of continuity
satisfied by the spin density, spin current density and spin torque
density:
\begin{eqnarray}\frac{\partial \rho_{\bf s}}{\partial t}
+\nabla\cdot {\bf J}^{\bf s}= T^{\bf s}.\end{eqnarray} This equation
and Eqs. (\ref{torque}) and (\ref{torque-p}) reveal that the matrix
$\hat{T}$ completely describes the spin precession caused by a total
effective magnetic field, which may contain both the Zeeman term and
the effective magnetic field originated from the spin-obit coupling.

A potential question is that whether the torque density matrix
$\hat{T}$ can also be rewritten as a divergence by solving the
equation $\hat{T}=-\nabla \cdot \hat{\bf J}_{T}$,\cite{Shi} although
the expression for $\hat{T}$ itself no longer explicitly contains
spatial derivatives $\nabla_{\bf R}$. This is not practicable,
because the condition $\hat{T}=-\nabla \cdot \hat{\bf J}_{T}$ is not
sufficient to uniquely determine the expression for $\hat{\bf
J}_{T}$ except in one-dimensional systems, since the $\hat{\bf
J}_{T}$ will have two or three unknown components in a 2D or
three-dimensional systems, but there is only one equation, which is
evidently not sufficient.\cite{footnotes} The physical meaning of
this conclusion is that the precession rate $\int
_{\Omega}\hat{T}dV$ in a volume ${\Omega}$ cannot be determined only
by the information on the boundary surface $\partial\Omega$, i.e.,
it cannot be expressed by a surface integral such as a current. In
contrast, the effect of the term $\nabla\cdot\hat{\bf
J}'_{\textmd{so}}$ can be expressed as a surface integral $\int
_{\partial\Omega}\hat{\bf J}'_{\textmd{so}}\cdot d{\bf S}$ if we
consider $(\partial/\partial t)\int_{\Omega} \hat{\rho}d^3{\bf R}$.
Furthermore, from definitions (\ref{C-operator}) and
(\ref{current-1}), we can find that there is a common factor
$\hat{p}_i-\hat{p}'_i=-i\hbar
\partial_{R_i}$ in the expression for $\hat{\bf J}'_{\textmd{so}}$,
which implies that the term $\int _{\partial\Omega}\hat{\bf
J}'_{\textmd{so}}\cdot d{\bf S}$ may still be nonvanishing when even
there are no any electrons inside the volume $\Omega$, i.e.,
$\hat{\rho}({\bf R})=0$ for all ${\bf R}\in \Omega$, because the
derivatives $\partial\hat{\rho}/\partial R_i$ on the boundary may
include a nonvanishing spin polarized part. So, it may give a
nonvanishing $\hat{\bf J}'_{\textmd{so}}$, and the
$(\partial/\partial t)\int_{\Omega}\hat {\rho}d^3{\bf R}\sim -\int
_{\partial\Omega}\hat{\bf J}'_{\textmd{so}}\cdot d{\bf S}$ may also
be nonvanishing; it turns out that the time evolution rate of the
total spin inside the volume is nonvanishing even when there are no
electrons. Therefore, the $\int _{\partial\Omega}\hat{\bf
J}'_{\textmd{so}}\cdot d{\bf S}$ can only be interpreted as another
kind of current flowing from the outside of the volume $\Omega$,
since in this case there are no spins (or electrons) inside the
volume; so, it is impossible to be explained as the spin precession
inside the volume.

\section{The influence of the Coulomb interaction and the scattering by nonmagnetic impurities}
Now, we consider the influence of the Coulomb interaction, which can
be written as
\begin{eqnarray}
{\bf H}_e=\frac{1}{2}\int d1d1'
\psi_{\sigma}^\dag(1)\psi_{\sigma'}^{\dag}(1')v(1-1')
\psi_{\sigma'}(1')\psi_{\sigma}(1),
\end{eqnarray}
where $v(1-1')$ is independent of spin. So, the total Hamiltonian is
${\bf H}+{\bf H}_e$, and Eq. (\ref{subtract-Dyson})
becomes\cite{Baym&Kadanoff, Baym-pr1962}
\begin{eqnarray}[\hat{G}_{0}^{-1}\hat{G}-\hat{G}\hat{G}_{0}^{-1}]_{\alpha\beta}(1,1')
=-i\hbar\int[v(1-\bar{2})-v(1'-\bar{2})]G_{\alpha\gamma,\gamma\beta}(1\bar{2}^-,1'\bar{2}^{+}),\label{eqofmotion}\end{eqnarray}
where $[\hat{G}_{0}^{-1}\hat{G}](1,1')=i\hbar(\partial/\partial
t)\hat{G}(1,1')-\hat{h}(1)\hat{G}(1,1')$ and $
[\hat{G}\hat{G}_{0}^{-1}](1,1')= -i\hbar(\partial/\partial
t')\hat{G}(1,1')-\hat{G}(1,1')\hat{h}(1');$
$G_{\alpha\gamma,\gamma'\beta'}(12,1'2')=[1/(i\hbar)^2]\langle
\emph{T}[\psi_{\alpha}(1)\psi_{\gamma}(2)\psi_{\gamma'}^{\dag}(2')\psi_{\beta'}^{\dag}(1')]\rangle$
is the two-particle Green's function, and a bar over variable
$(\bar{2})$ indicates that it is the integral variable,
$2^-\equiv(x_2,t_2-0^+).$ As $1'\rightarrow 1^+,$ the left-hand side
of Eq. (\ref{eqofmotion}) is $\hat{T}[\hat{\rho}]-(\partial
\hat{\rho}/\partial t)-\nabla \cdot\hat{\bf J},$ while the
right-hand side will vanish because $$
[v(1-\bar{2})-v(1'-\bar{2})]_{1'=1}=0.$$ Therefore, we again obtain
the Eq. (\ref{continuity}).

We take into account the scattering of some random nonmagnetic
impurities and suppose that the impurity potential has the form
$U({\bf x};\{{\bf r}_i\})=\sum_{i}u(x-{\bf r}_i),$ with ${\bf r}_i$
is the position of $i$th impurity and here $U(\{{\bf r}_i \})$ is a
function of the set $\{{\bf r}_i \}$. The additional term of the
second-quantized Hamiltonian ${\bf H}_{imp}(\{{\bf r}_i\})=\int
d1U(1;\{{\bf r}_i\})\psi^{\dag}_{\beta}(1)\psi_{\beta}(1)$. Now, the
interesting quantities become some quantities averaged over these
random impurity positions, e.g.,
$\overline{\hat{\rho}(1)}\equiv\int\cdots\int\prod_i(d^3{\bf
r}_i/\Omega)\hat{\rho}(1;\{{\bf r}_i\}),$ and $\overline{\hat{\bf
J}(1)}\equiv\int\cdots\int\prod_i(d^3{\bf r}_i/\Omega)\hat{\bf
J}(1;\{{\bf r}_i\}),$ etc., where $\Omega$ is the volume of the
system. So, the Eq. (\ref{subtract-Dyson}) becomes \begin{eqnarray}
\left[i\hbar\frac{\partial}{\partial
t}-e\phi(1)+i\hbar\frac{\partial}{\partial t'}+e\phi(1')\right
]\hat{G}(\{{\bf r}_i\})-\left[ \hat{h}+U(\{{\bf
r}_i\}),\hat{G}(\{{\bf r}_i\})\right ]=0.\end{eqnarray} Because $U$
is independent of spin, we have $$\lim_{1'\rightarrow 1^+}[U(\{{\bf
r}_i\}),\hat{G}(1,1';\{{\bf r}_i\})]=\lim_{1'\rightarrow
1^+}[U(1,\{{\bf r}_i\})-U(1',\{{\bf r}_i\})]\hat{G}(1,1';\{{\bf
r}_i\})=0.$$ Therefore, we get
\begin{eqnarray}
\frac{\partial \hat{\rho}({\bf x}t;\{{\bf r}_i\})}{\partial
t}+\nabla \cdot \hat{\bf J}({\bf x}t;\{{\bf r}_i\})=\hat{T}\left[
\hat{\rho}(\{{\bf
r}_i\})\right].\label{impurity-equation}\end{eqnarray} By taking the
ensemble average withe respect to $\{{\bf r}_i\}$ of both sides of
this equation, we have
\begin{eqnarray}\int\cdots\int\prod_i\frac{d^3{\bf r}_i}{\Omega}\left[
\frac{\partial \hat{\rho}({\bf x}t;\{{\bf r}_i\})}{\partial
t}+\nabla \cdot \hat{\bf J}({\bf x}t;\{{\bf r}_i\})\right ]
=\int\cdots\int\prod_i\frac{d^3{\bf r}_i}{\Omega}\hat{T}\left[
\hat{\rho}(\{{\bf r}_i\})\right],\nonumber\end{eqnarray} interchange
the ordering of the integrations and differentiations, we obtain
\begin{eqnarray}
\frac{\partial \overline{\hat{\rho} }}{\partial t}+\nabla \cdot
\overline{\hat{\bf J}}=\hat{T}\left[\overline{\hat{\rho}}\right
].\label{imp-continuity}\end{eqnarray}

Similarly, it is can also be proved that this equation still holds
if the Coulomb interaction and the impurity scattering exist
simultaneously. Furthermore, if we take the functional
differentiations with respect to the external fields on both sides
of this equation, we will obtain some exact identities connecting
the correlation functions of the spin current, spin torque, and the
usual particle current, which will be very similar to the usual
Ward-Takahashi identity even though the spin is not a conserved
quantity. So, if we use some approximation methods to calculate the
response of the spin current or spin torque to external
perturbations, then the self-consistency must be taken into account
in order to preserve these exact identities, just as in the problem
of the electron transport.\cite{Baym&Kadanoff,Baym-pr1962} In
addition, these expressions can also be applied to the finite
temperature systems, only if all averages are defined with respect
to a grand canonical ensemble, in which the spin-orbit coupling
terms will appear as a thermodynamic weighting factor
$\exp(-\hat{\bf H}_{\textmd{so}}/kT)$. Because the spin-orbit
coupling coefficients are often very small, so Green's functions and
the related observables may appear as spin polarized only if the
temperature is low enough such that $\hat{\bf H}_{\textmd{so}}/kT$
can not be ignored. Otherwise, this weighting factor will be almost
spin isotropic, so the phenomena due to the spin polarization will
be unobservable.

\section{Discussions and comparisons}
The expressions derived in Sec. II, which are expressed in terms of
density matrix, can also be given by the corresponding one-body
operators such that $\langle \hat{f} \rangle=\textmd{tr}\int d{\bf
x}d{\bf x}'\hat{f}({\bf x}',{\bf x})\hat{\rho}({\bf x},{\bf
x}')\equiv\textmd{Tr}(\hat{f}\hat{\rho})$, where $\hat{f}({\bf
x}',{\bf x})=\langle {\bf x}'|\hat{f}(\hat{\bf p},\hat{\bf x})|{\bf
x} \rangle$ represents a one-body operator; $\textmd{"Tr"}$ denotes
the trace over both the spin and coordinate variables (while the
$\textmd{"tr"}$ is the trace over spin). This form is more
convenient to compare with the conventional definition, since the
latter was usually given by a one-body operator.  To this end, we
need only to consider a special case $\hat{\rho}({\bf x},{\bf
x}')=\psi({\bf x})\psi^*({\bf x}')$ and obtain the following
relation:
\begin{eqnarray}&~&
[\hat{p}_{i_1}\cdots\hat{p}_{i_m}\hat{p}'_{j_{1}}\cdots\hat{p}'_{j_n}\hat{\rho}({\bf
x},{\bf x}')]_{{\bf x}'={\bf x}}=\langle\psi|
\hat{p}_{j_{1}}\cdots\hat{p}_{j_n}\delta(\hat{\bf x}-{\bf
x})\hat{p}_{i_1}\cdots\hat{p}_{i_m}|\psi\rangle\nonumber\\&=&\left
[\left(-i\hbar\widetilde{\partial}_{j_{1}}\right)\cdots
\left(-i\hbar\widetilde{\partial}_{j_{n}}\right)\psi({\bf x})\right
]^{*} \left [\left(-i\hbar\widetilde{\partial}_{i_{1}}\right)\cdots
\left(-i\hbar\widetilde{\partial}_{i_{m}}\right)\psi({\bf x})\right
],\label{one-body form}
\end{eqnarray}
where $\hat{\bf x}$ is the position operator and ${\bf x}$ is the
eigenvalue of $\hat{\bf x}$ [note that $\delta(\hat{\bf x}-{\bf
x})=|{\bf x}\rangle\langle {\bf x}|$ is the probability density
operator]. Here, the spin suffixes are ignored for simplicity. From
this relation, we have the following corresponding rule:
\begin{eqnarray}
\hat{p}_{i_1}\cdots\hat{p}_{i_m}\hat{p}'_{j_{1}}\cdots\hat{p}'_{j_n}\Longleftrightarrow
\hat{p}_{j_{1}}\cdots\hat{p}_{j_n}\delta(\hat{\bf x}-{\bf
x})\hat{p}_{i_1}\cdots\hat{p}_{i_m}.
\end{eqnarray}

According to this rule, we can obtain the corresponding one-body
operators of ${\bf J}_{0}$, ${\bf J}^{\bf s}_{0}$, ${\bf
J}_{\textmd{so}}$, ${\bf J}^{\bf s}_{\textmd{so}}$, $({\bf J}^{\bf
s}_{\textmd{so}})'$, and $T^{\bf s}$, which will be denoted by
$\hat{\bf j}_{0}$, $\hat{\bf j}^{\bf s}_{0}$, $\hat{\bf
j}_{\textmd{so}}$, $\hat{\bf j}^{\bf s}_{\textmd{so}}$, and
$\hat{t}^{\bf s}$, respectively. From definitions (\ref{current-j0})
and ${\bf J}_{0}=\textmd{tr}\hat{\bf J}_{0}$, we have
\begin{eqnarray}
\hat{\bf j}_{0}=\frac{1}{2m}\{ \hat{\bf p}, \delta(\hat{\bf x}-{\bf
x}) \}\end{eqnarray} and its mean value $\langle\psi| \hat{\bf
j}_{0}|\psi\rangle=(i\hbar/2m)\left[(\widetilde{\nabla}\psi)^{*}\psi-\psi^{*}\widetilde{\nabla}\psi
\right].$ Its contribution to spin current density is
\begin{eqnarray}
\hat{\bf j}^{\bf s}_{0}=\frac{1}{4}\left\{\hat{\bf s},\left\{\frac
{\hat{\bf p}}{m}, \delta(\hat{\bf x}-{\bf x})
\right\}\right\}=\frac{1}{2}\{\hat{\bf s},\hat{\bf
j}_{0}\}.\end{eqnarray} In order to express the one-body operators
$\hat{\bf j}_{\textmd{so}}$ and  $\hat{\bf j}^{\bf
s}_{\textmd{so}}$, we define
\begin{eqnarray}
\hat{d}^{(m)}_{i_1 \cdots i_{m}}(\hat{\bf p},\hat{\bf x})\equiv
\delta(\hat{\bf x}-{\bf x})\hat{p}_{i_1}\cdots
\hat{p}_{i_{m}}+\hat{p}_{i_{m}}\delta(\hat{\bf x}-{\bf
x})\hat{p}_{i_1}\cdots \hat{p}_{i_{m-1}}+\cdots +\hat{p}_{i_1}\cdots
\hat{p}_{i_{m}}\delta(\hat{\bf x}-{\bf x}).
\end{eqnarray} This is the one-body operator corresponding to definition (\ref{D-operator}).
According to definition (\ref{current_so}), ${\bf
J}_{\textmd{so}}=\textmd{tr}\hat{\bf J}_{\textmd{so}}$, and ${\bf
J}^{\bf s}_{\textmd{so}}=\textmd{tr}(\hat{\bf s}\hat{\bf
J}_{\textmd{so}})$, we obtain that the term of particle current
density $\hat{\bf j}_{\textmd{so}}$ can be given by
\begin{eqnarray}
(\hat{j}_{\textmd{so}})_{i}=\sum_{m}\hat{H}^{(m+1)}_{i,i_1\cdots
i_m} \hat{d}^{(m)}_{i_1 \cdots i_{m}}(\hat{\bf p},\hat{\bf
x}),\label{j_so operator}
\end{eqnarray}
while its corresponding spin current density can be given by
\begin{eqnarray}
(\hat{j}^{\bf s}_{\textmd{so}})_{i}=\sum_{m}\frac{1}{2}\{ \hat{\bf
s},\hat{H}^{(m+1)}_{i,i_1\cdots i_m}\} \hat{d}^{(m)}_{i_1 \cdots
i_{m}}(\hat{\bf p},\hat{\bf x})=\frac{1}{2}\{ \hat{\bf s},
(\hat{j}_{\textmd{so}})_{i}\}.\label{Js-operator}
\end{eqnarray} Similarly, let
\begin{eqnarray}
\hat{r}^{(m)}_{i_1 i_2\cdots
i_m}=\left\{\frac{\hat{p}_{i_1}}{2},\left\{\frac{\hat{p}_{i_2}}{2},\cdots,\left\{\frac{\hat{p}_{i_m}}{2},\delta(\hat{\bf
x}-{\bf x})\right\}\cdots\right\}\right\}
\end{eqnarray}
and
\begin{eqnarray}
\hat{c}^{(m)}_{i_1 i_2\cdots i_m}=[\hat{p}_{i_1}\cdots
\hat{p}_{i_m}, \delta(\hat{\bf x}-{\bf x})]+[\hat{p}_{i_1}\cdots
\hat{p}_{i_{m-1}},\hat{r}^{(1)}_{i_m}]+\cdots
+[\hat{p}_{i_1},\hat{r}^{(m-1)}_{i_2 \cdots i_m}].
\end{eqnarray}
From definitions (\ref{current-1}), (\ref{torque}),
(\ref{spin-current}) and (\ref{torque-s}) we have
\begin{eqnarray}
(\hat{j}^{\bf s}_{\textmd{so}})'_i=\sum_{m\geq
1}\frac{1}{4}[\hat{\bf s}, \hat{H}^{(m+1)}_{i,i_1\cdots
i_{m}}]\cdot\hat{c}^{(m)}_{i_1\cdots i_{m}},
\end{eqnarray}
\begin{eqnarray}\hat{t}^{\bf s}\equiv\sum_{n}
\frac{1}{i\hbar}\left[\hat{\bf s}, \hat{H}^{(n)}_{i_1i_2\cdots
i_{n}}\right]\hat{r}^{(n)}_{i_1 i_2 \cdots i_{n}}.\end{eqnarray}

In order to compare these expressions with the conventional
definition, we must first clarify a misleading statement in the
conventional definition. In recent literature, the spin current was
usually defined as $\hat{\bf J}^{\bf s}=\frac{1}{2}\{\hat{\bf
v},\hat{\bf s}\}.$ However, what we are actually concerned is not
the mean value of this operator $\langle\psi|\hat{\bf J}^{\bf
s}|\psi\rangle$, but the mean value of some kind of density of this
operator $\frac{1}{2}[\psi^{\dag}({\bf x})(\hat{\bf J}^{\bf
s}\psi({\bf x}))+(\hat{\bf J}^{\bf s}\psi({\bf x}))^{\dag}\psi({\bf
x})]=\langle\psi|\frac{1}{2}\{\hat{\bf J}^{\bf s},\delta(\hat{\bf
x}-{\bf x})\}|\psi\rangle$. Therefore, more precisely, the quantity
we actually defined is $\frac{1}{4}\{\{\hat{\bf v},\hat{\bf s}\},
\delta(\hat{\bf x}-{\bf x})\}$ rather than $\hat{\bf J}^{\bf s}$
itself and should be called a spin current density. Moreover, the
spin current, by analogy with any other kind of current, should be
defined as an integral $I^{\bf
s}[\textmd{A}]=\int_{\textmd{A}}j^{\bf s}_{i} d\textmd{A}_{i}$ of
the spin current density $j^{\bf s}_{i}$ over a surface
$\textmd{A}$, which is also not the $\frac{1}{2}\{\hat{\bf
v},\hat{\bf s}\}$ itself. Now, a puzzle that confronted us is that
there exist some different Hermitian combinations of the products of
$\hat{\bf v}$, $\delta(\hat{\bf x}-{\bf x})$ and $\hat{\bf s}$
because they do not commute in spin-orbit coupling systems. They all
have the same dimension and the same classical analogy (or classical
limit), it is very difficult to determine which one should be
interpreted as the proper spin current density just by this
intuitive method. For example, we can also define
$\frac{1}{4}\{\hat{\bf v},\{\delta(\hat{\bf x}-{\bf x}), \hat{\bf
s}\}\}$ or $\frac{1}{4}\{\{\hat{\bf v},\delta(\hat{\bf x}-{\bf
x})\}, \hat{\bf s}\}$ as a spin current density (only the first one
and the third one are equal owing to the fact that $\hat{\bf s}$ and
$\hat{\bf x}$ commute). From the classical picture of motion, the
physical meaning of the second one can be interpreted as the
velocity multiplied by the spin density; while third one is the
particle current density multiplied by spin. Moreover, if the
velocity includes higher power of momentum, i.e., it is a nonlinear
function of momentum, then there will be even more probable
expressions of the operator due to the noncommutative property of
the density (or position) and momentum. According to the results
given above, the spin current density must include two different
parts in order to satisfy the generalized continuity equation. The
first part is $\hat{\bf j}^{\bf s}_{0}+\hat{\bf j}^{\bf
s}_{\textmd{so}}$, it can be written as $\frac{1}{2}\{\hat{\bf s},
\hat{\bf j}_{0}+\hat{\bf j}_{\textmd{so}}\}$. However, the $\hat{\bf
j}_{\textmd{so}}$ cannot be simply written as $\hat{\bf
j}_{\textmd{so}}=\frac{1}{2}\{\frac{\partial
\hat{h}_{\textmd{so}}({\bf p})}{\partial {\bf p}}, \delta(\hat{\bf
x}-{\bf x})\}$ (only the $\hat{\bf j}_{0}$ does) but should be
written as a more complex expression given by Eq. (\ref{j_so
operator}). These two expressions are equal only if the velocity
operator $\frac{\partial \hat{H}({\bf p})}{\partial {\bf p}}$ is a
linear function of $\hat{\bf p}$ (correspondingly, the degree of the
full Hamiltonian must satisfies $\textmd{deg} \hat{H}(\hat{\bf
p})\leq 2$). The second part $(\hat{\bf j}^{\bf s}_{\textmd{so}})'$
is originated from the nonlinear spin-orbit coupling terms, it
appears only if $\textmd{deg} \hat{h}_{\textmd{so}}(\hat{\bf p})\geq
2$ and takes the form of commutator instead of anticommutator. From
the viewpoint of noncommutability of operators, this new term is
originated from the noncommutability of the density operator
$\delta({\hat{\bf x}-{\bf x}})$ and the $d\hat{\bf s}/dt$, because
the latter one also includes momentum operators in spin-orbit
coupling systems. This term does not give any additional
contribution to the particle current (or charge current) but may
contribute an additional surface integral term if we consider the
time rate of change of the total spin in a given volume.

As an example of these differences, we can consider a model with the
following Dresselhaus spin-orbit
coupling,\cite{Dresselhaus,Rashba-Sov,Dyakonov} in which
\begin{eqnarray}
\hat{h}_{\textmd{so}}=\gamma\sum_{i}\sigma_{i}\hat{p}_{i}(\hat{p}^2_{i+1}-\hat{p}^2_{i+2}),
~~~ (i=x,y,z;~~i+3\rightarrow i),
\end{eqnarray} where $\sigma_i$ are the Pauli matrices. After symmetrization, it can
be rewritten as $\hat{h}_{\textmd{so}}=\hat{H}^{(3)}_{ijk}\hat{p}_i
\hat{p}_j \hat{p}_k,$ where
$\hat{H}^{(3)}_{xyy}=\hat{H}^{(3)}_{yxy}=\hat{H}^{(3)}_{yyx}=-\hat{H}^{(3)}_{xzz}
=-\hat{H}^{(3)}_{zxz}=-\hat{H}^{(3)}_{zzx}=\frac{\gamma}{3}\hat{\sigma}_x,$
$\hat{H}^{(3)}_{yzz}=\hat{H}^{(3)}_{zyz}=\hat{H}^{(3)}_{zzy}=-\hat{H}^{(3)}_{yxx}
=-\hat{H}^{(3)}_{xyx}=-\hat{H}^{(3)}_{xxy}=\frac{\gamma}{3}\hat{\sigma}_y,$
$\hat{H}^{(3)}_{zxx}=\hat{H}^{(3)}_{xzx}=\hat{H}^{(3)}_{xxz}=-\hat{H}^{(3)}_{zyy}
=-\hat{H}^{(3)}_{yzy}=-\hat{H}^{(3)}_{yyz}=\frac{\gamma}{3}\hat{\sigma}_z,$
and all the other $\hat{H}^{(n)}_{i_1\cdots i_n}=0$. Because the
$\hat{\bf J}_0$ always has the same form for arbitrary model, so,
here we only give the expression for $\hat{\bf J}_{\textmd{so}}$,
$\hat{\bf J}'_{\textmd{so}}$ and $\hat{T}$. According to the
definitions (\ref{current_so}), (\ref{current-1}) and
(\ref{torque}), we have
\begin{eqnarray}
(\hat{J}_{\textmd{so}})_x
&=&\frac{\gamma}{6}\left(\{\hat{\sigma}_{x},(\hat{D}^{(2)}_{yy}-\hat{D}^{(2)}_{zz})\hat{\rho}\}-
\{\hat{\sigma}_{y},(\hat{D}^{(2)}_{xy}+\hat{D}^{(2)}_{yx})\hat{\rho}\}+
\{\hat{\sigma}_{z},(\hat{D}^{(2)}_{xz}+\hat{D}^{(2)}_{zx})\hat{\rho}\}\right)\nonumber\\
&=&\frac{\gamma}{6}(\{\hat{\sigma}_x,(\hat{p}^2_y+\hat{p}_y\hat{p}'_{y}+\hat{p}'^2_{y}
-\hat{p}^2_z-\hat{p}_z\hat{p}'_{z}-\hat{p}'^2_{z})\hat{\rho}\}
-\{\hat{\sigma}_y,(\hat{p}_x\hat{p}_y+\hat{p}_y\hat{p}_x+\hat{p}_x\hat{p}'_{y}\nonumber\\
&~&+\hat{p}_y\hat{p}'_{x}+\hat{p}'_{x}\hat{p}'_{y}
+\hat{p}'_{y}\hat{p}'_{x})\hat{\rho}
\}+\{\hat{\sigma}_z,(\hat{p}_x\hat{p}_z+\hat{p}_z\hat{p}_x+\hat{p}_x\hat{p}'_{z}+\hat{p}_z\hat{p}'_{x}+\hat{p}'_{x}\hat{p}'_{z}
+\hat{p}'_{z}\hat{p}'_{x})\hat{\rho}\})_{1'=1},
\end{eqnarray}
\begin{eqnarray}
(\hat{J}'_{\textmd{so}})_x
&=&\frac{\gamma}{12}\left([\hat{\sigma}_{x},(\hat{C}^{(2)}_{yy}-\hat{C}^{(2)}_{zz})\hat{\rho}]-
[\hat{\sigma}_{y},(\hat{C}^{(2)}_{xy}+\hat{C}^{(2)}_{yx})\hat{\rho}]+
[\hat{\sigma}_{z},(\hat{C}^{(2)}_{xz}+\hat{C}^{(2)}_{zx})\hat{\rho}]\right)_{1'=1}\nonumber\\
&=&\frac{\gamma}{8}\{[\hat{\sigma}_x,(\hat{p}^2_y-\hat{p}'^2_y
-\hat{p}^2_z+\hat{p}'^2_{z})\hat{\rho}]
-[\hat{\sigma}_y,(\hat{p}_x\hat{p}_y+\hat{p}_y\hat{p}_x+\hat{p}'_{x}\hat{p}'_{y}
+\hat{p}'_{y}\hat{p}'_{x})\hat{\rho}]\nonumber\\
&~&+[\hat{\sigma}_z,(\hat{p}_x\hat{p}_z+\hat{p}_z\hat{p}_x+\hat{p}'_{x}\hat{p}'_{z}+\hat{p}'_{z}\hat{p}'_{x})\hat{\rho}]\}_{1'=1},
\end{eqnarray}
The $y$ and $x$ components can be obtained by permutations of the
suffixes according the rule $x\rightarrow y, y\rightarrow z,
z\rightarrow x$ and $x\rightarrow z, z\rightarrow y, y\rightarrow
x$, respectively. The spin torque density is
\begin{eqnarray}\hat{T}[\hat{\rho}]=
\frac{\gamma}{i\hbar}\sum_{i}\left[\hat{\sigma}_{i},\frac{1}{8}(\hat{p}_{i}+\hat{p}'_{i})
\left((\hat{p}_{i+1}+\hat{p}'_{i+1})^2-(\hat{p}_{i+2}+\hat{p}'_{i+2})^2\right)\hat{\rho}\right]_{1'=1}.
\end{eqnarray}
For this model, ${\bf J}_{\textmd{so}}$ can not be simply given by
$\frac{1}{2}\{\hat{\bf v}_{\textmd{so}},\delta(\hat{\bf x}-{\bf
x})\}$ with $\hat{\bf v}_{\textmd{so}}=\partial
\hat{h}_{\textmd{so}}({\bf p})/\partial {\bf p}$, although their
results correspond to the same classical analogy. Because by the
latter intuitive definition all terms such as
$\hat{p}_{x}\hat{p}'_{x}$, $\hat{p}_{x}\hat{p}'_{y}$ etc. are
replaced by $\hat{p}_{x}\hat{p}_{x}$ (or
$\hat{p}'_{x}\hat{p}'_{x}$), and $\hat{p}_{x}\hat{p}_{y}$ (or
$\hat{p}'_{x}\hat{p}'_{y}$), respectively, so it obviously violates
the local equation of continuity. Moreover, the new term $\hat{\bf
J}'_{\textmd{so}}$, which is purely originated from the nonlinear
spin-orbit interaction and cannot be given by intuitive method, must
also be taken into account in order to satisfy the continuity
equation.

\section{Summary}
In conclusion, by defining a current and a torque density matrix, a
general equation of continuity satisfied by the matrixes of the
density, current density and torque density has been derived. This
equation holds in spin-orbit coupling systems as long as their
Hamiltonian can be expressed in terms of a power series in momentum.
Thereby, the universal expressions of the current density matrix and
torque density matrix have been uniquely determined. The current
density matrix can completely describe the particle (or charge)
current as well as the spin current in a unified form, while the
torque density matrix represents spin precession caused by a total
magnetic field, including a usual and an effective one. This
effective magnetic field is originated from the spin-orbit coupling
and depends on momentum. In contrast to the conventional intuitive
definition, it is found that if the spin-orbit coupling Hamiltonian
includes nonlinear term of momentum, then the definition of the
current density matrix should contain a new additional term, which
is purely originated from the nonlinear spin-orbit coupling and can
only contribute to the spin current. Moreover, if the degree of the
full Hamiltonian $\geq3$, then the conventional intuitive definition
of the current density must also be modified in order to satisfy the
local conservation law of number.

\begin{acknowledgments}
We are very thankful to Z. S. Ma and D. P. Li for their suggestions
and to Li Zhao, Rui Shen, Junren Shi and Gang Wu for helpful
discussions. This work was supported by the Research Fund for the
Doctoral Program of Higher Education.
\end{acknowledgments}
\appendix
\section{Derivation of EQ. (\ref{deverg-torque})}
Because (1) $\hat{p}_{i}$ is commute with $\hat{p}'_{i}$, and (2)
${H}^{(n)}_{i_1\cdots i_{n}}$ is symmetric under the interchange of
its suffixes, we have
\begin{eqnarray}\hat{H}^{(m+1)}_{i_1\cdots i_{m+1}}\left(
\hat{p}_{i_1}\cdots \hat{p}_{i_{m+1}} -\hat{R}^{(m+1)}_{i_1\cdots
i_{m+1}}\right)
\hat{G}(1,1')=\frac{1}{2}\left(\hat{p}_{i}-\hat{p}'_{i}\right)\hat{H}^{(m+1)}_{i,i_1\cdots
i_m}\hat{F}^{(m)}_{i_1\cdots i_{m}} \hat{G}(1,1'),\end{eqnarray}
\begin{eqnarray}\hat{H}^{(m+1)}_{i_1\cdots i_{m+1}}\left(
\hat{p}'_{i_1}\cdots \hat{p}'_{i_{m+1}} -\hat{R}^{(m+1)}_{i_1\cdots
i_{m+1}}\right)
\hat{G}(1,1')=-\frac{1}{2}\left(\hat{p}_{i}-\hat{p}'_{i}\right)\hat{H}^{(m+1)}_{i,i_1\cdots
i_m}\hat{F'}^{(m)}_{i_1\cdots i_m} \hat{G}(1,1'),\end{eqnarray}
where
\begin{eqnarray}\hat{F}^{(m)}_{i_1\cdots i_m}=\hat{p}_{i_1}\cdots \hat{p}_{i_{m}}+
\hat{p}_{i_1}\cdots \hat{p}_{i_{m-1}}\hat{R}^{(1)}_{i_m}+\cdots+
\hat{R}^{(m)}_{i_1\cdots i_m},\end{eqnarray}
\begin{eqnarray}\hat{F'}^{(m)}_{i_1\cdots i_m}=\hat{p}'_{i_1}\cdots \hat{p}'_{i_{m}}+
\hat{p}'_{i_1}\cdots\hat{p}_{i_{m-1}}\hat{R}^{(1)}_{i_m}+\cdots+\hat{R}^{(m)}_{i_1\cdots
i_m}.\end{eqnarray} Therefore, we get
\begin{eqnarray}\hat{H}^{(m+1)}_{i_1\cdots i_{m+1}}\left\{
\hat{p}_{i_1}\cdots \hat{p}_{i_{m+1}},
\hat{G}\right\}-2\hat{H}^{(m+1)}_{i_1\cdots
i_{m+1}}\hat{R}^{(m+1)}_{i_1\cdots i_{m+1}}
\hat{G}=\frac{1}{2}\left(\hat{p}_{i}-\hat{p}'_{i}\right)\hat{H}^{(m+1)}_{i,i_1\cdots
i_m}\hat{C}^{(m)}_{i_1\cdots i_m} \hat{G},\label{A7}\end{eqnarray}
where
\begin{eqnarray}\hat{C}^{(m)}_{i_1\cdots i_m}&=&\hat{F}^{(m)}_{i_1\cdots i_m}-\hat{F'}^{(m)}_{i_1\cdots
i_m}\nonumber\\
&=& \hat{p}_{i_1}\cdots
\hat{p}_{i_{m}}+\hat{p}_{i_1}\cdots\hat{p}_{i_{m-1}}\hat{R}^{(1)}_{i_{m}}\cdots
+\hat{p}_{i_1}\hat{R}^{(m-1)}_{i_2\cdots i_{m}}\nonumber\\&-&
\hat{p}'_{i_1}\cdots
\hat{p}'_{i_{m}}-\hat{p}'_{i_1}\cdots\hat{p}'_{i_{m-1}}\hat{R}^{(1)}_{i_{m}}
\cdots -\hat{p}'_{i_1}\hat{R}^{(m-1)}_{i_2\cdots
i_{m}}.\end{eqnarray} Similarly, we have
\begin{eqnarray}\left\{
\hat{p}_{i_1}\cdots \hat{p}_{i_{m+1}},
\hat{G}\right\}\hat{H}^{(m+1)}_{i_1\cdots
i_{m+1}}-2\hat{R}^{(m+1)}_{i_1\cdots
i_{m+1}}\hat{G}\hat{H}^{(m+1)}_{i_1\cdots i_{m+1}}
=\frac{1}{2}\left(\hat{p}_{i}-\hat{p}'_{i}\right)\hat{C}^{(m)}_{i_1\cdots
i_m} \hat{G}\hat{H}^{(m+1)}_{i,i_1\cdots
i_m}.\label{A8}\end{eqnarray} Adding Eq. (\ref{A7}) and (\ref{A8}),
we get
\begin{eqnarray}\frac{1}{2}\left[\hat{H}^{(m+1)}_{i_1\cdots
i_{m+1}},\left\{ \hat{p}_{i_1}\cdots \hat{p}_{i_{m+1}},
\hat{G}\right\}\right]=\frac{1}{4}\left(\hat{p}_{i}-\hat{p}'_{i}\right)\left
[\hat{H}^{(m+1)}_{i,i_1\cdots i_m},\hat{C}^{(m)}_{i_1\cdots
i_m}\hat{G}\right ]+\left[\hat{H}^{(m+1)}_{i_1\cdots
i_{m+1}},\hat{R}^{(m+1)}_{i_1 \cdots
i_{m+1}}\hat{G}\right].\end{eqnarray} By taking the limit as
$1'\rightarrow 1^+$, we obtain the Eq. (\ref{deverg-torque}).


\begin{thebibliography}{10}
\bibitem{footnotes} {This ununiqueness can also be verified by considering some concrete
examples. e.g., in a 2D circle of radius $R$, the vector field ${\bf
V}({\bf x})=f(r)(-y/r, x/r)$ defined inside this circle obviously
satisfies $\nabla\cdot{\bf V}({\bf x})=0$, here $f(r)$ is arbitrary
differentiable function satisfying $f(0)=f(R)=0$, $r$ is the
distance between $(x,y)$ and $(0,0)$, so, the solution of
$\nabla\cdot {\bf J}_T({\bf x})=T$ is unnuique when even the
boundary condition is given, it can be arbitrarily altered by ${\bf
J}_T+{\bf V}$. For 3D system, the solution of the equation
$\nabla\cdot{\bf J}_T({\bf x})=T$ can be altered by ${\bf
J}_T+\nabla\times {\bf F}$, where ${\bf F}$ is an arbitrary vector
field.}
\end{thebibliography}
\end{document}